# Intracrystalline inclusions within single crystalline hosts: from biomineralization to bio-inspired crystal growth

Eva Weber and Boaz Pokroy[*]

Department of Materials Science and Engineering and the Russell Berrie Nanotechnology Institute, Technion Israel Institute of Technology, Haifa 32000, Israel

*bpokroy@tx.technion.ac.il*

**Abstract**

Many crystals in nature exhibit fascinating mechanical, optical, magnetic and other characteristics. One of the reasons for this phenomenon has to do with the presence of specific organic molecules that are tightly associated with the mineral. Over the years, some organic crystals have been found to be located within the lattices of their single-crystalline biogenic hosts. A number of questions remain unanswered: for example, how do these molecules become incorporated and what is their function? In this review we survey the gradual refinement of the above mentioned finding in biogenic crystals, with the object of tracing the acquisition of our fundamental knowledge in this field during the last 50 years. We highlight the progress made in understanding the function and significance of this intracrystalline organic matter, from the earliest observations of this phenomenon in a biological system to the highly promising recent achievements in bio-inspired material synthesis, where intracrystalline molecules have been used in many studies to synthesize numerous synthetic nanohybrid composites with fascinating new properties.

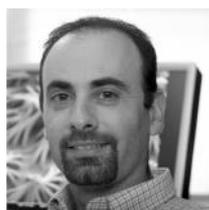

Boaz Pokroy is an associate professor in the Department of Materials Science and Engineering at the Technion–Israel Institute of Technology, Israel, where he earned all of his degrees. Previously, he was a postdoctoral fellow and Fulbright Scholar in the laboratory of Prof. Joanna Aizenberg at the School of Engineering and Applied Sciences at Harvard University and Bell Labs. Pokroy's research focuses on biomineralization and bio-inspired surface engineering. He studies the structure of biominerals on the atomic, nano-, and mesoscales using state-of-the-art high-resolution characterization techniques such as high-resolution synchrotron diffraction and aberration-corrected transmission electron microscopy. Based on the strategies that organisms use to produce natural materials, his lab also develops novel bio-inspired materials, such as semiconductors whose bandgap can be tuned by the incorporation of intracrystalline biological molecules; controlling the short-range order of nano-amorphous materials; and fabrication of superhydrophobic/superoleophobic surfaces for various applications.

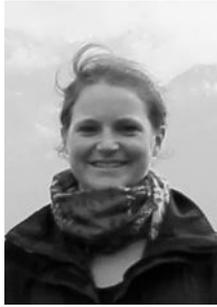
Eva Weber studied human- and molecular biology at University of Saarland, Germany with focus on plant physiology. After graduation she joint the field of biomineralization and received her PhD degree from the University of Saarland. Her PhD work was carried out at the Institute of New Materials at Saarland University in the group of PD Dr. Habil. Ingrid Weiss. After receiving a Minerva scholarship, she is conducting her postdoctoral research in the group Assoc. Prof. Boaz Pokroy, at the Technion, Israel Institute of Technology, Israel. Her research interests encompass the interdisciplinary field of biomineralization with focus on the interaction of biomolecules and inorganic matter. Further, her interests cover plant biology and how mineral formation in plants is influenced by abiotic stress.

Table of Content:

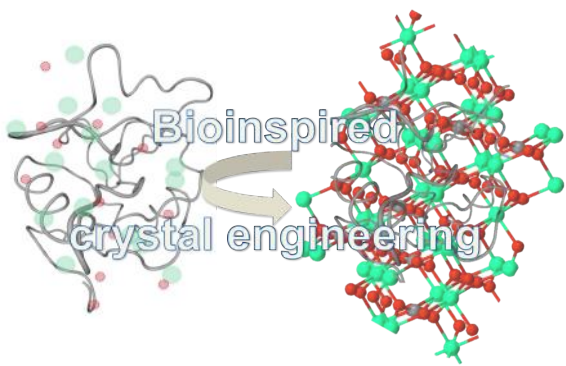

A review of the inclusion of organic matter within single crystalline hosts: from biogenic minerals to bio-inspired nano-hybrid single crystal composites.

# 1. Introduction

Biominerals, the crystals produced by organisms in the process of biomineralization, have long been a source of fascination to scientists because of their intriguing shapes and morphologies and their superior material properties.[1-7] Biogenic minerals are comprised of several hierarchical levels over different length scales. As an outcome of this organization, these materials demonstrate higher crack resistance as compared to their non-biogenic counterparts. Organisms use biominerals for a variety of functions such as for protection,[1] ion-storage[8] and light scattering[9], as a few examples. In order to obtain these remarkable functions, organisms need to precisely direct mineral growth. In contrast to non-biological inorganic crystals, biominerals are hybrid nanocomposites that contain an inorganic and an organic phase. The latter is found to be present in a broad range of concentrations from as low as 0.05 and as high as 40 %.[2] Today, one knows that the organic phase is one of the major factors in directing mineral growth.[10]

The organic phase of a biomineral can be divided into two main types according to the localization of its organic molecules. Whereas "intercrystalline" molecules are localized at the surfaces of the crystals, *i.e.*, at grain boundaries between single crystals, "intracrystalline" organic matter refers to organic molecules distributed within a single crystal host. Together, the two types cover a very large area of research. As the former type has been extensively discussed elsewhere,[11-19] this review is focused mainly on the latter type, *i.e.*, the intracrystalline incorporation of organic molecules.

We review the findings leading to our current knowledge about intracrystalline molecules. We want to emphasize that the vast majority of reports concerning intracrystalline molecules are related to calcium carbonate both in the realm of biogenic crystals as well as bio-inspired. To the best of our knowledge there are no reports yet on the characterization of intracrystalline molecules in other biomineral systems other than the ones mentioned herein. Based on this, the review starts with the first observations made for intracrystalline molecules in nature, followed by the finding of their superior material properties that could be correlated to the presence of intracrystalline organic molecules. Subsequently, we discuss the current knowledge of the microstructure of biominerals and what distinguishes them from their non-biogenic counterpart. In addition, we examine the technical attempts to visualize the biomolecules within the hosting crystal lattice structure by utilizing state-of-the-art characterization techniques; after which we provide a short survey of intracrystalline molecules that have already been characterized. Finally, we consider how the acquired knowledge can potentially be put to use in the realm of synthetic bio-inspired materials engineering as summarized in the section "Bio-inspired Crystal Synthesis".

Despite intensive research and proliferating knowledge in the field of biomineralization, many questions still remain to be answered:
How do such macromolecules succeed to become incorporated into dense inorganic crystalline hosts? Why do they not disrupt the crystallinity of the host? What are the criteria governing whether or not a specific molecule will become incorporated into a specific crystal? What is the biological function of these intracrystalline molecules?
These are important issues, since biominerals are complex nanocomposite materials and elucidation of their internal structure therefore requires a combination of techniques as will be demonstrated later on in this review.

## 2. First observations

The first studies reporting the presence and distribution of intracrystalline molecules were already published as long as half a century ago and were based on observations from various sea organisms. The formation of mollusk shells was studied by Norimitsu Watabe, who paid particular attention to the distribution of organic matter in biogenic minerals. His decalcification experiments revealed an intercrystalline and an intracrystalline matrix in the nacre of the bivalve species *Pinctada martensii, Elliptio complanatus* and *Crassostrea virginica* and indicated that the intracrystalline molecules are arranged in a sheet-like substructure.[22]

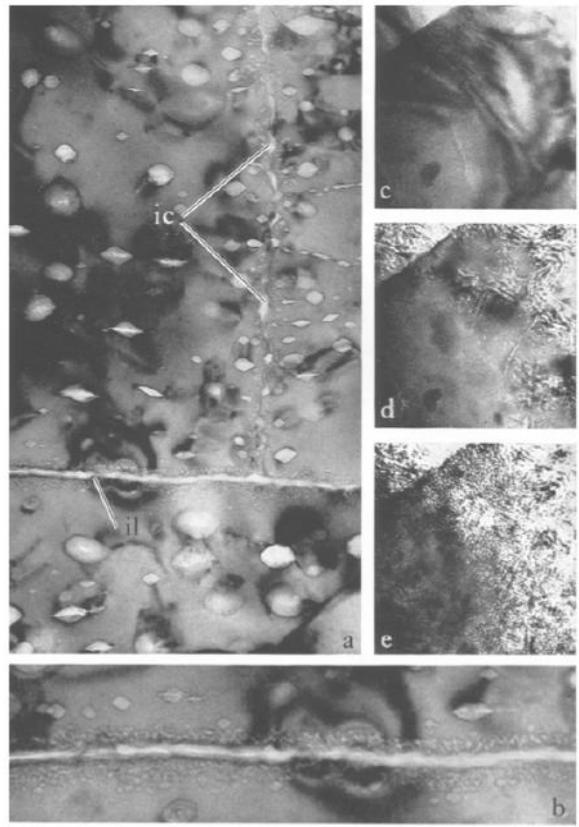

**Fig. 1**. Transmission electron micrographs of ion-beam micromilled nacre samples (a,b). Atypical inclusions are concentrated close to boundaries of the nacre platelets. These structures are distinct from the electron beam-induced damage obtained from non-biogenic aragonite (c-e) (from Towe and Thompson[23], Fig. 4).

In 1972, Towe and Thompson observed "bubbly" and "frothy" structures within nacre tablets upon examining the *Mytilus* nacre by transmission electron microscopy (TEM) (Fig. 1). These features were found to be concentrated at interlamellar locations (Fig. 1a, 'il') and also at intercrystalline areas (Fig. 1a, 'ic'), and to occur specifically in the aragonitic layer rather than in the calcitic prismatic layers. To determine whether these structures had been created as a result of beam damage or were real, non-artifactual inclusions, Towe and Thompson investigated non-biogenic aragonite in a comparable manner at different electron-beam intensities (Fig. 1c,d). The results of those experiments enabled the authors to conclude that the biogenic samples possess distinct features not seen in the non-biogenic counterpart mineral. They further concluded that the observed intracrystalline features in biogenic aragonite originate from "trapped water and trapped organic material".[23]

## *2.1* **Biogenic crystals fracture as single crystals**

Biominerals exhibit a wide variety of unique biocomposite material properties,[24-27] however the distinctiveness of their fracture properties in particular was useful in highlighting the presence and influence of single-crystal inclusions in these materials. To this end, in 1969 the attention of scientists was attracted by the spines of the sea urchin because of their peculiar behavior in fracture experiments. Whereas non-biogenic calcite was known to fracture preferentially along its {104} cleavage planes (Fig. 2a,b), the sea urchin's spines, composed of single crystals of magnesium calcite underwent conchoidal fracture during cracking[28] and exhibited surfaces very similar to that of amorphous glass (Fig. 2b,c).[29] This behavior was unexpected as sea urchin spines had previously been reported to diffract as single crystals.[28,30]

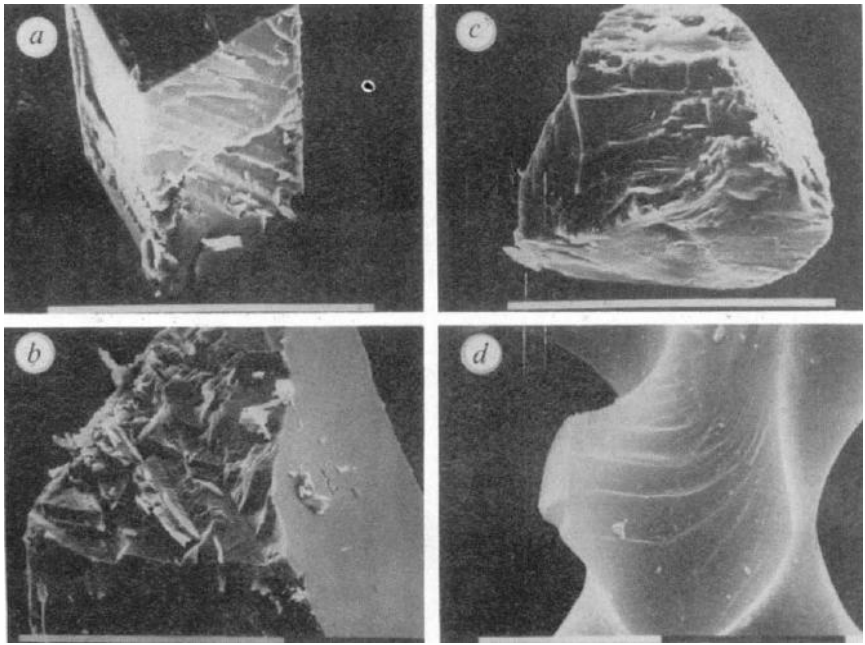

**Fig. 2**. Scanning electron microscopic images of cleaved calcite in the absence (a,b) and in the presence (c,d) of organic compounds (from Berman *et al*.,[29] Fig. 3).

In an attempt to understand this observation, Berman *et al*. precipitated synthetic calcite in the presence of organic molecules extracted from single-crystalline mineralized parts of the sea urchin. This procedure yielded crystals exhibiting fracture properties reminiscent of those seen in the biogenic crystals. This observation clearly indicated that the material properties of the single-crystalline biominerals differed from those of their synthetic counterparts, and that the alteration was probably due to their incorporation of organic matter.[29] The authors further suggested that the intracrystalline organic matter located at mosaic boundaries and specific crystal planes[3,29,31] influences the cleavage behavior of the biogenic crystals relative to that of their non-biogenic counterparts.[31] Based on these findings, close attention was paid to deciphering of the crystal microstructure.

## *2.2* **Microstructure of biogenic crystals**

Not only the crystal morphology but also the microstructure of biogenic crystals turned out to be distinct from non-biogenic minerals. An intriguing observation was made when analyzing sea urchin spines on a synchrotron beamline.[31,33] Compared to their synthetic counterparts, the biogenic samples exhibited a decrease in coherence length and an increase in peak widths and anisotropy. Similar observations were made when intracrystalline molecules extracted from sea urchin spines were allowed to become incorporated *in vitro* into growing single crystals of calcite. These results demonstrated that biogenic biomolecules are indeed able to become integrated into the microstructure, thereby altering it *in vitro*.[34]

Further proof that a decrease in X-ray coherence length and angular spreads are widespread phenomena in biogenic crystals was provided by a comparison of single crystals from sea urchin (*Paracentrotus*) with those from mollusks (*Atrina*) and foraminifer (*Patellina*). The crystal texture was found to be distinct between species, most likely as a result of the way in which the organic phase interacts with the inorganic host. For example, XRD experiments revealed that in the case of *Paracentrotus*, macromolecules seem to be aligned mainly along the calcite *c*-axis, whereas in the prisms of *Atrina* the biomolecules are arranged perpendicular to the calcite *c*-axis.[33] The extent to which organisms are able to control their crystal microstructure and morphology was further elegantly demonstrated by the example of a reduction in crystal symmetry observed in calcite spicules, a phenomenon not observed in control experiments with synthetic calcite.[35]

In their microstructural studies utilizing XRD, Berman *et al.* assumed that the observed broadening of the diffraction peaks originates entirely from crystallite size (coherence length). Based on this assumption they used the Scherrer equation to compare the anisotropic broadening of different biogenic calcites to that of synthetic calcite samples as a control. This comparison yielded good correlation between the anisotropic coherence lengths and the global morphology of the crystals.[31] Later, Pokroy and Zolotoyabko performed a more detailed XRD microstructural analysis on biogenic calcite, taking into account the effect not only of coherence length but also of microstrain fluctuations on the diffraction peak broadening.[36] Moreover, in that study the authors compared the evolution of these two characteristics as a function of isochronous annealing within the same sample. Such annealing disrupted the organic/inorganic interfaces, changing the microstructure. It turned out that the grain size was strongly reduced, with an accompanying increase in the microstrain fluctuations. This pronounced anisotropic change in microstructure clearly identified the specific planes on which the intracrystalline organic molecules had been situated prior to annealing. These observations were very different from the opposite behavior usually observed in conventional materials, and turned out to be a landmark phenomenon for any crystal in which organic molecules are incorporated. Similar observations were also made in biogenic aragonite.[37] For further details on how such experiments are conducted see Pokroy *et al.*, 2007.[37]

## *2.3* **Lattice distortions in biogenic crystals: a widespread phenomenon**

In the years that followed, Pokroy and Zolotoyabko conducted a comparative study of biogenic calcium carbonate using synchrotron-based high-resolution powder XRD combined with the Rietveld refinement method.[38] The latter was applied to determine lattice parameters of biogenic crystals (aragonite in this case) with the highest precision possible. Based on the values obtained and with geological aragonite crystals used as a control, the biogenic aragonitic lattice was shown to be anisotropically distorted. Relative to control, a maximum distortion of ~0.1% was found along the crystallographic c-direction in biogenic aragonite of the *Acanthocardia tuberculata* bivalve mollusk shell[39] and calcite.[40] The amounts of inorganic impurities in the aragonitic lattice were too low (by an order of magnitude) to account for the relatively high distortions observed. These distortions could be fully relaxed via a rather mild heat treatment at a temperature as low as 140°C. In view of the low temperature that allowed for relaxation of the lattice distortions, combined with the low levels of impurities, the authors concluded that the phenomenon was due to the incorporation of intracrystalline biomolecules. Pokroy and Zolotoyabko further demonstrated similar lattice distortions in a wide range of aragonitic biogenic crystals collected from different classes and habitats (fresh and salt water as well as land; see Fig. 3) and exhibiting various microstructures (nacre, prismatic and crossed lamellar).[37,41] By means of neutron diffraction, Pokroy and Zolotoyabko further showed that owing to these incorporated molecules, not only is the unit cell of aragonite distorted anisotropically but also the structure is distorted in terms of bond lengths and angles.[42] Enlargement of lattice parameters for biogenic aragonite has also been reported for the shell of the mollusk *Tapes decussatus*.[43] Further, macrostrain along all crystallographic axes was found in aragonite of the marine bivalve *Anomia simplex*.[44] In addition, results from a comparative study in corals (*Favia* and *Desmophyllum)*[45] were in good agreement with the findings of Pokroy and co-workers.[37,41]

The observation that lattice parameters are anisotropically changed in biogenic calcium carbonate relative to the non-biogenic mineral indicates that organic molecules are incorporated into the crystal lattice. This conclusion was strengthened by the use of additional methods such as small-angle X-ray scattering (SAXS). In this case, annealing led to an increase in the X-ray scattering contrast owing to destruction of the organic/inorganic interfaces. The experimental results indicated that the (001) planes are the specific planes to which the organic molecule species preferentially adhere.[46]

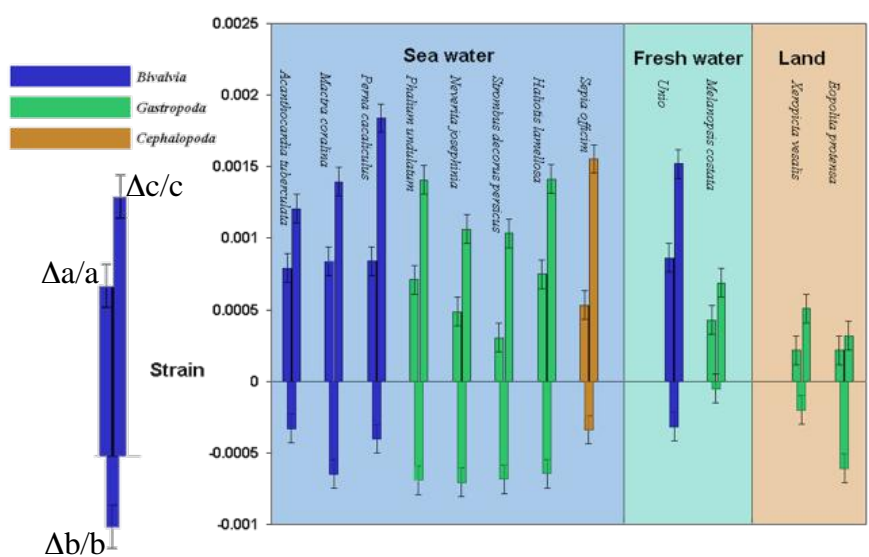

**Fig. 3.** Lattice distortions determined for sea and land shells. Distortions are calculated on the basis of values obtained in geological aragonite (from Pokroy *et al.*,[41] Fig.1).

## *2.4* Direct visualization of organic macromolecules within biogenic crystal hosts

Imaging of the organic minor phase within an inorganic crystalline host is extremely challenging, especially as the organic phase is mainly comprised of carbon and the calcium carbonate matrix is also carbon-rich. During the crystal growth process the surfaces of such hybrid nanocomposites can indeed be imaged by surface techniques such as atomic force microscopy (AFM), but we do not necessarily obtain valid information about how the organics are arranged within the crystal below the surface.[47,48] Electron microscopy was shown to be an appropriate tool for investigating biominerals with high resolution.[49] Because organic molecule species are known to be highly sensitive to beam damage, the data obtained must be evaluated carefully. Despite such methodical limitations, some studies have beautifully revealed the distribution of intracrystalline molecules. One such study was performed by means of annular dark-field scanning transmission microscopy (ADF-STEM) coupled with 3D-data reconstruction. Using this method, Li and coworkers mapped the organic matter in single calcitic prisms of *Atrina rigida* and revealed their internal anisotropic distribution.[50] A similar approach was employed by Younis *et al.* to image the individual distribution of organic molecules within single platelets of aragonite from the nacreous layer of the green mussel (*Perna canaliculus*), using high-angle annular dark-field scanning electron microscopy (HAADF-STEM) in the tomography mode and electron tomography reconstruction. Their results elegantly showed that the observed organic patches are not scattered uniformly within a single crystal but rather aligned along the (001) plane (see Fig. 4).[51] Both studies indicated that biomolecules are differentially incorporated into the lattice of the crystal hosts of calcite and aragonite. In the latter example their observations are in good agreement with information we obtained from our powder XRD measurements, namely that lattice distortions appear most prominently along the *c*-direction of the crystal host. However, Younis and colleagues observed a "depletion zone" containing fewer organic patches but a more brush-formed distribution of organics close to the intercrystalline organic sheets separating neighboring nacre platelets (Fig. 4).[51] These observed "depletion zones" seem to be structurally distinct from the residual platelets and might also differ among species. As described earlier in this review, Towe and Thompson in 1972 (see Fig. 1 above) observed an accumulation of inclusions in the area close to the organic layers.[23] Non-homogeneous nanosized inclusions in mollusk shells were also detected by TEM.[52,53] In these latter studies, measurements obtained by Electron-Energy-Loss Spectroscopy (EELS) revealed a higher content of carbon within these specific inclusions, indicating that these areas contain biopolymers.[53]

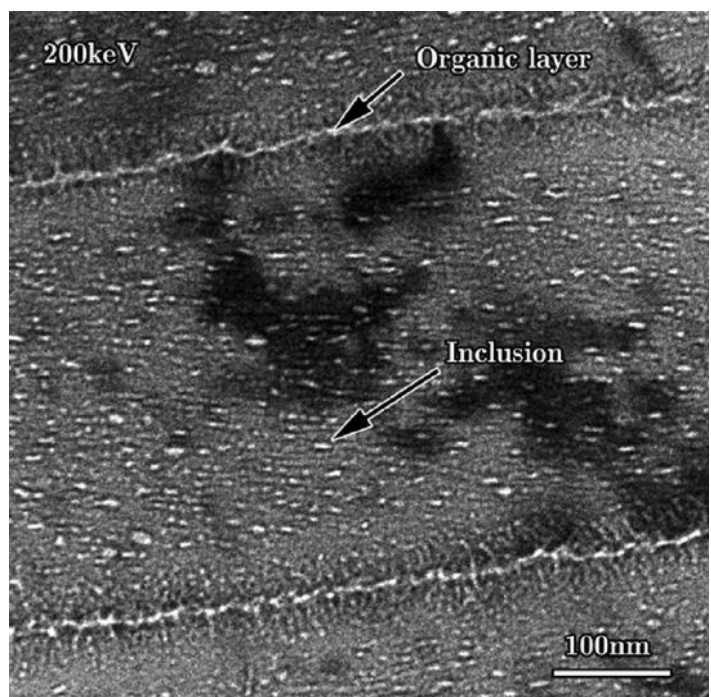

**Fig. 4**. Intracrystalline organic patches appear as bright spots within nacre lamellae when imaged by TEM (bright field mode) (from Younis *et al.,*[51] Fig. 5).

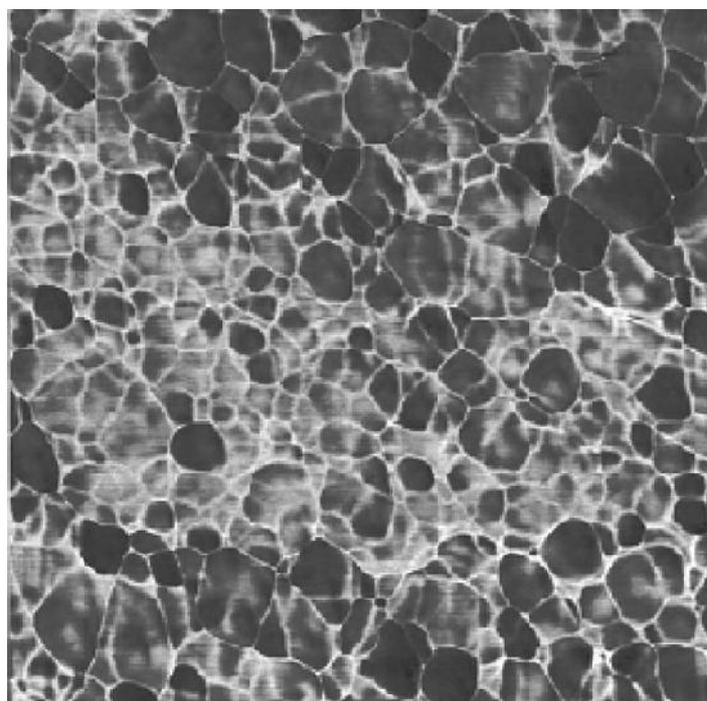

Fig. 5. AFM image of nacre indicating parts of the intracrystalline organic network of a nacre tablet from *Pinctada maxima* (from Rousseau *et al.*,[54] Fig: 4). Image size 1 μm × 1μm.

Using AFM and TEM, intracrystalline molecules were visualized as 45-nm "vesicles" encapsulating oriented nanogranules/nanotablets of calcium carbonate (Fig. 5). Rousseau *et al.* showed that the intracrystalline organic fraction exhibits an internal crystalline structure that seems to diffract as single crystals.[54] A comparable nanostructure was observed within the prismatic layer of the red abalone shell.[55]

Whereas powder XRD reveals information about the long-range order in minerals, Raman spectroscopy, infrared spectroscopy, X-ray absorption near-edge structure spectroscopy (XANES), atom-probe tomography, and X-ray photoelectron emission microscopy are powerful tools that provide information on chemical bonding as well as on the short-range order of molecules in biogenic minerals[56-58] and their corresponding precursor phases.[59] One example is the peptide, studied by Metzler and coworkers. The authors showed not only that the short-range order is altered in the biocomposite relative to non-biogenic calcite but also that asp2 protein interacts via a chemical bond with the calcite phase.[57] This had previously been demonstrated only for aragonite (based on high-resolution neutron-diffraction experiments)[60] and for amorphous calcium carbonate.[59] An additional powerful technique used to study the local environment of biomolecules in minerals is solid-state nuclear magnetic resonance. Following this route, Ben-Shir *et al.* successfully characterized organic/inorganic interfaces in the mollusk shell of *Perna canaliculus* and distinguished two types of molecular species: one was exposed and shown to interact with bio-organics whereas the other seemed to be less exposed and to interact with water and bicarbonate molecules.[61]

### 2.5 Examples of intracrystalline biomolecules

A wide range of literature can be found discussing biomolecules related to biomineralization processes.[62-65] However, the characterization and study of the composition of intracrystalline molecules began about 30 years ago. Subsequently, extraction methods were established to separate intercrystalline from intracrystalline molecular species. To obtain the intracrystalline fraction exclusively, the extraction has to be carefully conducted in order to remove all of the adsorbed and weakly bound biomolecules. A widely accepted method encompasses NaOCl treatment with subsequent dissolution of the crystalline phase using EDTA or HCl prior to analysis. (For more information on advantages and disadvantage of EDTA versus HCl extraction see Albeck *et al.*[66]) However, as the use of EDTA was found to potentially produce artifacts[67] and to interfere with upstream procedures, each protocol has to be adapted according to individual requirements.[68-71] In an alternative method an ion-exchange resin is used to dissolve the crystalline phase.[67] (For further discussion of extraction methods and classification of molecular species see Pereira-Mourie *et al.*[72])

The major compound identified after extraction, however, is of proteinogenic origin.[66] Although species-specific differences are found, intracrystalline and intercrystalline molecules share some common features based on their amino acid composition. These include a high content of acidic amino acids, some amino acids that are secondarily modified by sugar residues,[66,69,73-75] and chitin fibers.[70] In addition, XANES indicated the presence of sulfated sugar residues.[76] Based on macromolecules derived from sea urchin and mollusks, Albeck and coworkers demonstrated that the glycosylation state of proteins may alter the mineral-protein interaction.[77]

With the aim of gaining a better understanding of organic-inorganic interaction, single molecules were investigated in several studies. Antibody labeling of caspartin, a protein from the mollusk *Pinna nobilis*, clearly demonstrated that specific molecules are located both inter- and intraprismatically.[78,79] Caspartin, when used for *in-vitro* experiments, was shown to induce twinning in synthetic calcite.[80] Both caspartin and calprismin were extracted after extended bleach treatments, indicating their intimate attachment to the crystal phase.[79] Although these two proteins have some similarities their glycosylation states differ, as shown by the finding that calprismin is modified by a sugar residue whereas caspartin is not.[79] This finding might indicate that glycosylation does not necessarily direct protein-mineral interaction in this case, but might rather serve to fine-tune protein solubility.

A number of studies have been devoted to SM50, an intracrystalline protein from the spicule matrix of sea urchin.[81-84] In an additional study, Zhang and coworkers hypothesized that SM50 possesses several functional domains encompassing a mineral-recognition region, a self-assembly region, and a molecular-spacer region.[85] A recent study of the influence of recombinant SM50 protein domains on calcium carbonate mineralization showed that the distinct domains interact diversely with the mineral phase and favor protein aggregation, while also affecting early-stage formation of calcium carbonate.[86] To gain a better understanding of the protein's function, however, it would be worth investigating whether its peptides and recombinant protein derivatives are incorporated into the crystal lattice, and which domain is needed to mediate this process. This would also throw more light on recent observations in the bivalve *Pinctada margaritifera*, in which the authors describe an orientational gradient and splitting phenomena that are most likely related to intracrystalline molecular species.[87]

Little is known about structural properties of proteins related to biomineralization processes. In a bioinformatics analysis published in 2012 J.S. Evans discusses structure- and sequence-specific features of biomineralization proteins associated with aragonite.[88] This approach revealed, as defined by the author, that intracrystalline proteins (water-soluble fraction) possess fewer intrinsically disordered regions than framework- or pearl-associated proteins.[88] This study shows that bioinformatics analysis should be carried out in addition to experimental data evaluation and might allow one in the future to predict the behavior of macromolecules during interaction with a growing mineral phase.

In addition to the extensively studied calcium carbonate system, reports about the study of intracrystalline molecules in other biominerals are extremely scarce. Some reports on intracrystalline biomolecules refer to the mineral calcium oxalate found in urinary stones[89-91] and plant minerals.[92,93]

## 3. Bio-inspired crystal synthesis

Several approaches have been tried in an attempt to mimic concepts of crystal growth in nature[94-98] and further understand the phenomenon of intracrystalline incorporation. Such synthetic routes have yielded striking structures with extraordinary properties. These strategies are also aimed at determining specific conditions for optimizing crystal growth with respect to their specific function.

It is well accepted today that organisms often utilize different organic matrices as templates for biomineralization processes. Various organic compounds have been identified, for example, in mollusk shells, such as chitin fibers, silk-like proteins, and proteins enriched in acidic amino acids and modified with sugar residues. About 15 years ago a model of a decalcified mollusk shell, devised on the basis of cryo-TEM studies, was used to demonstrate how biomacromolecules are arranged in the living organism. β-chitin fibers are highly ordered and serve as a framework, whereas silk proteins serve as a hydrated gel to generate a confined space for the growing crystal.[99] Proteins in β-sheet conformation were shown to be ordered to some extent and linked to the chitin framework, and are most likely involved in directing crystal growth.[99,100] In the latter cited study the authors showed an epitaxial match between the structure of β-chitin and the *ab* plane of aragonite. Later it was shown that pure epitaxy can induce aragonite orientation.[101,102] Based on these findings, polymers attract increasing attention and were subsequently utilized for templating crystal growth. When synthesizing polymers it is possible to control a large variety of their properties (size, charge, compounds) such that various conditions for mineralization experiments can be examined. For this reason, we will focus on the polymer-based growth in the next paragraph followed by the discussion of how molecules can be used to alter and improve crystal properties in the second part of this section.

## 3.1 Polymer-based growth

Over the last few years, polymer-assisted methods have been found to be a powerful means of directing crystal growth in artificial systems.[103-106] One strategy, inspired by nature, is based on hydrogels or hydrogel-like structures. By creating a confined space in which crystals can grow and which simultaneously serves as an ion reservoir, Estroff and Li demonstrated that several types of hydrogels are applicable in crystal growth.[107] As reviewed in 2012, agarose, charged polysaccharides, gelatin, polyacrylamide, silica gel and silk were used as matrix for crystal growth.[107]

To the best of our knowledge, induction of strain by intracrystalline molecules grown in a gel matrix has not been reported. However, Li and coworkers, by growing calcite within an agarose network, elegantly solved the distribution of an agarose network incorporated into a single crystal of calcite. Their study was accomplished by HAADF-STEM and a 3D reconstruction series. The observed distribution of agarose fibers within single calcite crystals was rather random, but no disordered crystal lattice was detected. Subsequent heat treatment led to the formation of cavities within the single crystal in a lace-like net of continuous fiber-like structures.[108] Another interesting experiment was the growth of single crystals of the hen egg white protein lysozyme within a silica gel matrix. Although silica was indeed shown to be incorporated into the crystal structure, significant changes in crystal structure could not be detected.[109]

In other experiments latex particles were used as templates for mineralization leading either to the formation of a porous surface of single crystalline calcite[110] or were incorporated into the crystalline structure of zinc oxide. In the latter case the observed lattice strain could be released by a mild annealing treatment. No correlation was found between polymer concentration and strain intensity, in contrast to the concentration-dependent increase in strain observed when a biopolymer was incorporated into the crystal lattice of calcite.[111]

Furthermore, block copolymers that can be tuned were found to direct mineral growth.[106,112-114] The block copolymer $PSPMA_{30}-PDPA_{47}$ was successfully incorporated into the crystal lattice of calcite. These inclusions appeared as 20-nm micelles, mainly adsorbed to the (104) facets of the crystal host. Up to 13wt% polymer was incorporated within these synthetic single crystals of calcite,[114] whereas relative to biological systems up to 20% of the crystal volume was found to consist of organic cavities.[32] Additional studies report the incorporation of magnetite nanoparticles[115] and functionalized polystyrene[116] into single crystals of calcite. Nanoparticles of various appearances were occluded in a zeolitic imidazolate framework, which provided them with catalytic, magnetic and optical properties.[117]

## 3.2 Tuning crystal properties

Drawing inspiration from nature, several approaches have been employed to create biocomposite materials and to tune the properties of inorganic crystals. Moreover, bio-inspired approaches are frequently applied to incorporate organic[118-121] or inorganic compounds into the crystal host,[115,122,123] as reviewed above. Kahr *et al.* utilized the first approach to systematically study the incorporation of organic dyes into crystals of potassium acid phthalate ($C_8H_5KO_4$), thereby introducing a striking change in their optical properties (see Fig. 6).[119] Furthermore, this model system made it possible to study the interaction of dyes with individual crystal planes on a molecular level.[119] For a comprehensive review of the incorporation of organic molecular species into the crystal host see Kahr and Gurney 2001 and references therein.[120]

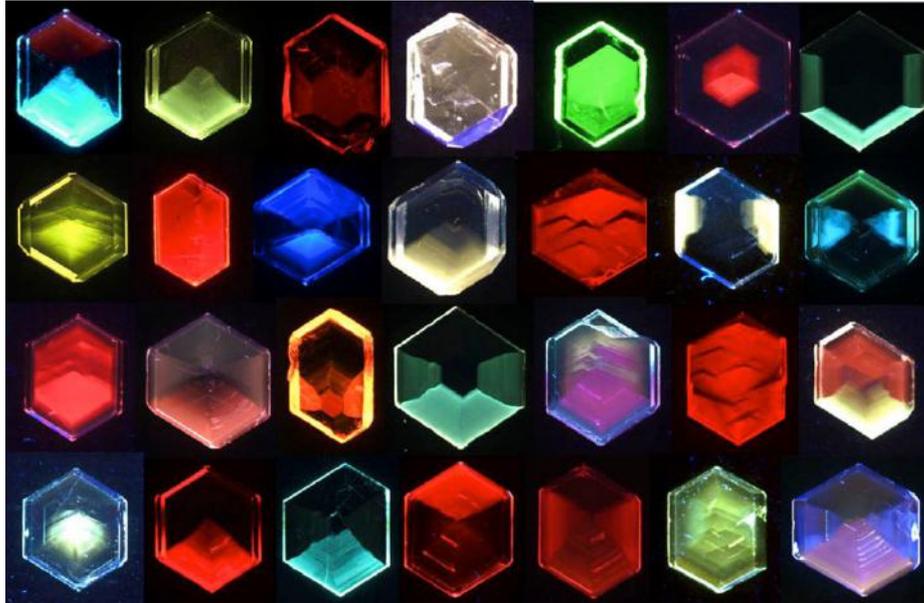

**Fig. 6.** Incorporation of organic molecules into the crystal of potassium acid phthalate $C_8H_5KO_4$ leads to changes in their optical properties (from Kahr *et al.*,[119] Fig. 2).

The role of amino acids in mineralization was recently scrutinized,[124] and in a basic study all of the amino acids were screened for their incorporation into the crystal lattice of calcite.[121] It was shown that especially aspartic acid and cysteine become incorporated at the highest levels and induce significant lattice distortions in the crystal host. Many other amino acids also showed significant incorporation in and distortion of the calcite host lattice. In light of this pioneering study,[121] Brif *et al.* showed that specific amino acids can also be incorporated into other crystalline hosts, particularly ZnO. The amino acids incorporated into ZnO induce lattice distortions that are also accompanied by a modification of the band gap of ZnO (Fig. 7a).[125,126]

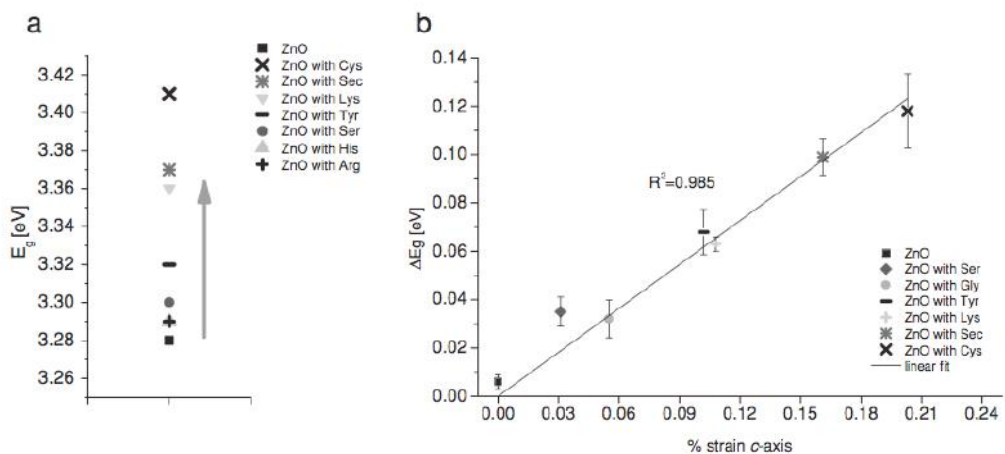

**Fig. 7**. Changes in band-gap values after incorporation of amino acids into zinc oxide (a). Correlation between band-gap energy change and *c*-axis strain determined for ZnO-containing amino acids (b) (From Brif *et al.*,[125] Fig. 3).

New optical and magnetic properties were introduced into the crystal structure of calcite by the incorporation of gold-oxide,[122] magnetite-oxide[115,123] and zinc oxide[123] nanoparticles. At almost the same time Liu and coworkers succeeded in conferring paramagnetic properties on calcite grown in an agarose gel, via the incorporation of $Fe_2O_3$ nanoparticles into the calcite or by dyeing of the calcite via incorporation of gold nanoparticles. In both cases the crystal lattice of calcite was not significantly disrupted and the desired result was not achievable when the calcite was grown in a solution-based system.[127]

Engineered peptides fused to proteins with optical properties can serve as an efficient tool not only for binding to inorganic materials but also for monitoring biomimetic processes *in vitro*.[128,129] Although peptides are studied mainly in terms of their surface-binding properties, it would be worth determining the extent to which these constructs might be incorporated into the crystal structure and alter the property of the material. Peptides are highly suitable for studying the influence of single molecules on crystals and for allowing us to acquire deeper knowledge about the required molecular sequences and physical properties.[130]

A highly desirable outcome in this field of study would be to predict the sequences and the 3D structures of intracrystalline protein or other molecule inclusions required for their incorporation within single crystalline hosts. However, in light of the above review, it is clear that various parameters have to be taken into account, in terms not only of macromolecular properties, sequences, structure and molecule density but also of the microenvironment and properties where the relevant processes are conducted.

## 4. Conclusions and outlook

Although by now we have gained a substantial amount of fundamental knowledge about the existence of such intracrystalline molecules, many questions still remain. A major question yet to be addressed concerns how specifically these macromolecules interact with the mineral phase. Although this topic is still in its infancy, the first attempts to decipher the local environment of intracrystalline macromolecules are very promising. Less information is available, however, about the specific function of intracrystalline proteins, and we still lack the requisite detailed knowledge about the role of single molecules and their impact on mineral formation. Some light has already been shed on these questions following the successful initial attempts at crystal engineering, which have opened new routes towards the achievement of new and improved materials. There is good reason for optimism that the bio-inspired approach will yet yield exciting new results.

## Acknowledgements

Funding from the European Research Council under the European Union's Seventh Framework Program (FP/2007–2013)/ERC Grant Agreement n° [336077] is greatly appreciated. We also thank the Minerva foundation for financial support.